# Giant odd-parity magnetoresistance from proximity-induced topological states

Tomoki Hotta[1], Le Duc Anh[1,2*], Takahiro Chiba[3,4], Yohei Kota[5], and Masaaki Tanaka[1,2,6*]

[1]*Department of Electrical Engineering and Information Systems, The University of Tokyo*
[2]*Center for Spintronics Research Network (CSRN), The University of Tokyo*
[3] *Department of Information Science and Technology, Graduate School of Science and Engineering, Yamagata University, Japan*
[4]*Department of Applied Physics, Graduate School of Engineering, Tohoku University*
[5]*National Institute of Technology, Fukushima College*
[6]*Institute for Nano Quantum Information Electronics (NanoQuine)*
*Corresponding authors: anh@cryst.t.u-tokyo.ac.jp, masaaki@ee.t.u-tokyo.ac.jp

**Magnetoresistance typically exhibits even symmetry with respect to the magnetic field, owing to time-reversal symmetry (TRS) as dictated by Onsager's reciprocity relations. However, in certain systems where TRS is broken, magnetoresistance may acquire an odd component with respect to the magnetic field—referred to as odd-parity magnetoresistance (OMR). To date, reported OMR values have been modest, usually restricted to a few tens of percent even under high magnetic fields (e.g., 14 T). Here, we report the discovery of a giant OMR reaching up to 1,150% under a relatively low field of 1 T in a heterostructure composed of 3 nm-thick α-Sn and a ferromagnetic semiconductor, (In,Fe)Sb. Although α-Sn in this thickness range is a trivial narrow-gap semiconductor, analysis of Shubnikov–de Haas oscillations combined with *ab initio* calculations reveals the emergence of tilted topological surface states, induced via magnetic proximity from the (In,Fe)Sb layer. The observed OMR behavior is well explained by a Boltzmann transport model assuming the presence of oppositely tilted Weyl cones in the α-Sn band structure. Our findings not only shed new light on the physics of OMR but also suggest promising avenues for its application in electronic and spintronic devices, such as ultra-sensitive magnetic sensors.**

Achieving large magnetoresistance (MR) is crucial for the advancement of information processing and sensing technologies. In narrow-gap semiconductors and semimetals with small Fermi surfaces, large and non-saturating MR has been observed and extensively studied[1,2]. Recently, considerable attention has been paid to materials with linear band dispersion, such as Dirac and Weyl semimetals, which have emerged from the broader field of topological materials. This linear dispersion can be viewed as a special case of systems with narrow band gaps or small Fermi surfaces. When time-reversal symmetry (TRS) is broken in a Dirac system, the Dirac point splits into a pair of Weyl nodes with opposite spin polarization. Weyl semimetals are well known for their unconventional magnetotransport phenomena, including the chiral anomaly[3] and Weyl orbits[4]. In addition, certain systems exhibit odd-parity magnetoresistance (OMR)[5–8], characterized by an antisymmetric MR component with respect to magnetic field reversal—an effect that generally contradicts Onsager's reciprocity relations. However, OMR can arise in materials with additional broken symmetries, such as magnetic Weyl semimetals (e.g., $Eu_2Ir_2O_7$[5], $SmCo_5$ and $Cd_2Os_2O_7$[6], $Co_3Sn_2S_2$[7], and $EuB_6$[8]), Dirac semimetals doped with magnetic impurities (e.g., $(Cd,Zn,Mn)_3As_2$[9]), and heterostructures exhibiting magnetic proximity effects (e.g., InAs/(Ga,Fe)Sb[10], $Fe_3GeTe_2$/graphite/$Fe_3GeTe_2$[11], and bilayer graphene/$Cr_2Ge_2Te_6$[12]). The magnitude of OMR in these systems is typically modest, often reaching only a few percent per Tesla (see Supplementary Table S1).

In this study, we report a giant odd-parity magnetoresistance (OMR) effect reaching up to 1,150% at 1 T—more than three orders of magnitude greater than any previously reported values. This remarkable enhancement is observed in an epitaxially grown bilayer consisting of a narrow-gap semiconductor α-Sn (3 nm) and a ferromagnetic semiconductor (In,Fe)Sb on an InSb (001) substrate, as illustrated in Fig. 1a. Details of the sample growth are provided in the *Methods* section. As illustrated in Fig. 1b, α-Sn is a diamond-type crystal structure with a lattice constant of 0.6489 nm and can be epitaxially grown on InSb (lattice constant: 0.6479 nm) using molecular beam epitaxy (MBE)[13-16]. When the thickness of α-Sn exceeds 10 nm on an InSb substrate, it exhibits a topological Dirac semimetal phase[17,18], characterized by a pair of Dirac cones arising from band inversion, accompanied by surface Fermi arcs connecting the nodes. In contrast, ultrathin α-Sn films with thicknesses below 20 monolayers (approximately 3 nm) revert to a topologically trivial narrow-gap semiconductor due to the resolution of band inversion[13]. However, when such a thin α-Sn film is interfaced with a ferromagnetic layer (e.g., Fe), the magnetic proximity effect can reinstate band inversion, thereby restoring the nontrivial topological phase[19]. In our system, we utilize (In,Fe)Sb as the ferromagnetic layer to break time-reversal symmetry (TRS). (In,Fe)Sb is a ferromagnetic semiconductor that exhibits ferromagnetism above room temperature [20,21] and shares the zinc-blende crystal structure of InSb, with a closely matched lattice constant of 0.6444 nm [22]. This structural compatibility allows for high-quality epitaxial growth of α-Sn on (In,Fe)Sb and facilitates the induction of magnetism in α-Sn through the magnetic proximity effect.

Crystal structure of the α-Sn/(In,Fe)Sb heterostructure was characterized by cross-sectional scanning transmission electron microscopy (STEM), as shown in Fig. 1c. The STEM lattice image confirms a zinc-blende crystal structure without any visible second phase of the (In,Fe)Sb layer, and a diamond crystal structure without any other precipitates of the α-Sn layer. Note that the top surface of the Sn layer is naturally oxidized, and the thickness of α-Sn layer is about 3 nm (~ 18 ML). The α-Sn layer thickness is determined by comparing the STEM lattice image and energy dispersive X-ray (EDX) mapping as shown in Fig. 1d. Room-temperature ferromagnetism of the (In,Fe)Sb layer was confirmed by magnetic circular dichroism (MCD) spectroscopy and superconducting quantum interference device (SQUID) magnetometry (See Supplementary note 1 and Fig. S1). Prior to the transport measurements, we have patterned the sample into Hall bar devices with a size of 200 μm × 50 μm by the standard photolithography and Ar ion etching technique.

We first present the observation of the giant OMR in Fig. 2. When measured at 2 K with an in-plane magnetic field applied parallel to the current direction (***B*** // ***I***) and a constant current of 10 μA, the α-Sn/(In,Fe)Sb heterostructure exhibits a large magnetoresistance accompanied by an antisymmetric component against the magnetic field. To rule out the contribution of the Hall effect, which would be largest when ***B*** is perpendicular to the plane, we apply the magnetic field in the film plane, parallel or antiparallel to the sample current (Fig. 2a,b). The OMR is extracted from the raw MR data $R(B)$ by $R_{odd}(B) = [R(B) - R(-B)]/2$, and the OMR ratio is estimated by $R_{odd}(B) / R(0\ T)$, where $R(0\ T) = 0.68\ \Omega$, as shown in Fig. 2c,d. The OMR ratio monotonically proportional to the magnetic field, even though it is not a simple linear function. At 1 T, the OMR ratio exceeds about 1,000%. When we rotated the magnetic field from the in-plane to the perpendicular direction to the thin film, the OMR changes approximately as a cosine function of the angle $\theta$, which is defined as the angle between the magnetic field ***B*** and the current ***I*** (Fig. 2e). The OMR disappears when ***B*** is perpendicular to the film, which excludes the Hall effect as a possible origin. It reaches maximum when ***B*** // ***I***, with a magnitude of 1,150% at 1 T, 2 K. This giant OMR is larger by three orders of magnitude than any other OMR reported thus far[5–12]. Moreover, the appearance of OMR when ***B*** // ***I*** suggest an origin related to the magnetic chiral effect[3], a characteristics of magnetic Dirac/Weyl semimetals.

To understand the origin of this giant OMR, we characterize the band structure of α-Sn by magnetotransport measurements. It is noteworthy that the (In,Fe)Sb layer is relatively insulating, especially at low temperatures[20], and thus most of the current flows in the top α-Sn layer (see Supplementary note 2 and Figure S2). When applying a magnetic field ***B*** up to 14 T perpendicular to the film plane, the sample exhibits a huge magnetoresistance about 200,000% at 2 K and clear Shubnikov-de Haas (SdH) oscillations (Fig 3a). To analyze the SdH oscillations following the Lifshitz-Kosevich theory[23,24], we extract the oscillatory component against the inverse magnetic field of the conduction data by subtracting a polynomial fit (Fig. 3b). Fast Fourier transformation (FFT) spectra of the oscillations (Fig. 3c) indicates that there is a single Fermi surface with a small frequency of 10.0 T. By fitting the theory to the temperature dependence of the FFT peak intensity (inset of Fig. 3c), we estimate an effective mass of $0.16m_0$ of the conduction electrons. A fan plot of the oscillation extrema at 2 K (Fig. 3d) indicates a Berry phase of 0.65. Considering that the Berry phase is zero for a quadratic band dispersion and 0.5 for a linear band dispersion[24], the observed SdH oscillation clearly indicates that the Fermi level is located in a linear dispersion of the α-Sn thin film. Furthermore, angle dependence of the SdH frequency (Fig. 3e) indicates that the observed SdH oscillation comes from a two-dimensional Fermi surface of the α-Sn layer (detail of discussion is described in Method section).

The two-dimensional Fermi surface of the α-Sn thin film observed in the SdH oscillation can be well explained by our first-principles calculations. As previously mentioned, a bulk α-Sn under compressive strain shows an inverted band structure where the *s*-orbital lies below the *p*-orbitals, thus making α-Sn topologically non-trivial, as confirmed in the calculated result shown in Fig. 3f. When α-Sn is 18-ML-thick (2.92 nm), quantum confinement resolves the band inversion, pushing the *s*-orbital just above the *p*-orbitals and forming linear band dispersion with a very small gap (Fig. 3g). This topological phase transition due to quantum confinement is experimentally confirmed in our previous study[13]. To calculate the band structure of the α-Sn (3 nm)/(In,Fe)Sb bilayer in this study, we utilize a model where Fe atoms are distributed on the top layer of a 18-ML-thick α-Sn (See details of the calculations in Methods section). As shown in Fig. 3h, hybridization with the Fe *d*-orbitals largely alters the band structure of the 18-ML-thick α-Sn and a nearly-zero-gap linear dispersion is formed around –0.25 eV both in the *Γ-X* and *Γ-Y* directions. This feature likely corresponds to the linear dispersion observed in the SdH oscillation analysis, as α-Sn grown on InSb tends to be lightly *p*-type doped due to In diffusion[13]. Here, *X* and *Y* are on the Brillouin zone boundary in ***k***-space along the $k_x$ and $k_x$ directions, respectively (See Supplementary Fig. S4). Note that the band dispersion is slightly different between the *Γ-X* and *Γ-Y* directions, which suggests that the linear dispersion is tilted in the *XY* plane. This tilted linear band dispersion plays a crucial role in inducing the giant OMR, as will be discussed below.

To further investigate the physics of OMR, we investigate angular dependence of the OMR when ***B*** is rotated in all three directions (Fig. 4, left). We set the magnetic field at 0.2 T and measured $R_{odd}$ while rotating ***B***. The *x*-axis is defined to be parallel to the current direction of the Hall bar, and the *z*-axis is normal to the sample surface (inset of Fig. 4). The labels '*xz*', '*yz*', '*xy*' correspond to the ***B*** rotation in the *x-z*, *y-z*, and *x-y* planes, respectively. The zero value of angle is defined when ***B*** is parallel to ***I*** (pointing along the *x* axis) in the *xz* and *xy* rotation data, and to the *y* axis in the *yz* rotation data. Note that when the magnetic field is stronger than 0.2 T, nonlinear response starts to appear in OMR as seen in Fig. 2c,d, which complicates the situation. Here, to focus on the data in the linear response, we have set the magnetic field at 0.2 T. As shown in Fig. 4a, the OMR is observed in all three directions of ***B*** rotation and most prominent around 0 degree and 180 degree in the *xz* rotation, 105 degree and 285 degree in the *yz* rotation, and around 35 degree and 215 degree in the *xy* rotation.

A previous theoretical study[25] based on the Boltzmann transport model showed that topological semimetals with pairs of Weyl cones tilting oppositely to each other (inset of the right figure of Fig.4b) exhibits a current component that is proportional to the magnetic field, which leads to the appearance of OMR. We notice that the *ab initio* calculated band structure of the α-Sn/(In,Fe)Sb heterostructure suggests this tilting dispersion to be likely. According to the theory, the odd-parity component of the current density under an electric field ***E*** and a magnetic field $\boldsymbol{B} = (B_x, B_y, B_z)$ is given as

$$j_x^{(1)} \propto 61 w_x E B_x + 7 w_y E z B_y , \qquad (1)$$

where $w_x$ and $w_y$ are constants expressing the tilt of Weyl cones in α-Sn. Derivation of eq. (1) is described in Methods section. When we set $\frac{w_y}{w_x} = 4.29$, the theoretically predicted current density well reproduces the angular dependence of the experimentally observed OMR, as shown in Fig. 4a and b. The fitted parameter $\frac{w_y}{w_x} = 4.29$ means that the tilt condition is anisotropic between the $k_x$ and $k_y$ directions. This anisotropy is consistent with the anisotropy of band structure in the *Γ-X* and *Γ-Y* directions seen in the calculated the band structure (Fig. 3h). This good agreement between the experimental result (Fig. 4a) and theoretical calculation (Fig. 4b) strongly supports the tilted Weyl cones model as the origin of the giant OMR observed in the α-Sn/(In,Fe)Sb bilayer heterostructure.

Finally, we comment on the potential applications of this giant OMR effect. Because the resistance of the α-Sn/(In,Fe)Sb bilayer varies both proportionally and antisymmetrically with the applied magnetic field, the OMR can be exploited for use in versatile magnetic field sensors, analogous to Hall sensors but offering significantly higher sensitivity. Given a resistance change of 1,150% at 1 T, this system holds great promise for realizing magnetic-field sensors with sensitivities up to an order of magnitude higher than those of commercially available ultrahigh-sensitivity Hall sensors based on InSb (see Methods). Unlike Hall sensors, whose sensitivity is fundamentally limited by the carrier mobility, the OMR-based sensors demonstrated in this work may offer further room for improvement—contingent on a comprehensive understanding of the physical mechanism.

In summary, we have successfully fabricated an epitaxial α-Sn thin film / (In,Fe)Sb heterostructure and observed a giant odd-parity magnetoresistance (OMR) reaching 1,150% at 1 T and 2 K. This extraordinary OMR is well accounted for by the theoretical model of tilted Weyl cones in topological Weyl semimetals. The observed OMR ratio exceeds previously reported values by more than two orders of magnitude, highlighting its strong potential for electronic applications, particularly in the development of high-sensitivity magnetic field sensors.

**Methods**

*Sample preparation.* α-Sn thin film / (In,Fe)Sb heterostructure was grown on an InSb(001) substrate by MBE. *In situ* reflection high energy electron diffraction (RHEED) patterns taken along the [110] direction during the MBE growth of each layer are shown in Supplementary Fig. S3. The substrate was affixed to a molybdenum block using indium and loaded into the MBE chamber. After desorbing the native oxide layer of the InSb substrate by annealing at a substrate temperature ($T_{sub}$) of 420 °C for 15 minutes under an $Sb_4$ flux, a 100-nm-thick InSb buffer layer was grown at $T_{sub}$ = 400°C under $Sb_4$ overpressure. Then $T_{sub}$ was cooled to 320°C under an $Sb_4$ flux and a 25-nm-thick InSb layer was grown. The $Sb_4$ flux was reduced such that the InSb buffer shows a $c(8 \times 2)$ pattern (Supplementary Fig. S3), which is essential for the subsequent α-Sn growth[13]. After the buffer layer growth, $T_{sub}$ was cooled to 250°C[19] in the absence of $Sb_4$ flux and a 7-nm-thick (In,Fe)Sb layer was grown with the same $Sb_4$ flux. During the (In,Fe)Sb layer growth, the $c(8 \times 2)$ reconstruction RHEED pattern was maintained (Supplementary Fig. S3). Then $T_{sub}$ was further cooled to 0°C, which is needed to grow the α-Sn thin film while preventing structural phase transition to β-Sn. A 5-nm-thick α-Sn layer was grown at 1 nm/min with a streaky RHEED pattern. Such a thin α-Sn layer is thermally stable up to around 160°C[26], which is also confirmed by the fact that the diamond crystal structure of α-Sn was maintained after heating it to the melting point of indium (~156°C) upon detachment of the sample following the growth procedure.

*Analysis of SdH oscillations.* To investigate the dimensionality of the Fermi surface, we measure the angle dependence of the frequencies of the SdH oscillations from the derivative of the Hall resistance against the magnetic field ***B***. Moreover, when rotating ***B*** from the perpendicular direction to an in-plane direction of the thin film, the FFT frequency shifts following $1/\cos\theta$, as shown in Fig. 3e. This indicates that the observed Fermi surface is two-dimensional. Considering that (In,Fe)Sb is insulating at low temperature[20], this two-dimensional Fermi surface must belong to the α-Sn layer. Note that another frequency that is independent of the ***B*** direction appears when the tilting angle of ***B*** is larger than 50 degrees (Fig. 3e), which reflects a three-dimensional band component. This three-dimensional Fermi surface possibly corresponds to the bulk InSb.

*First principles calculations.* The first-principles density functional theory calculation was performed by Vienna ab initio simulation package (VASP) with the Perdew-Burke-Ernzerhof generalized gradient approximation (GGA-PBE)[27,28]. The ion-electron interaction is described by projected augmented wave (PAW) potentials. Spin–orbit coupling was taken into consideration in the self-consistent calculations of the electronic band structure. The cutoff energy of the plane wave basis was fixed to 500 eV. We adopted a simple model, α-Sn/Fe slab geometry, instead of the experimental system represented by α-Sn/(In,Fe)Sb slab geometry, to capture the essence of the magnetic proximity effect on a α-Sn thin film (the detail of the model is described in Supplementary note 4). The Brillouin zone integrations in bulk and slab geometry calculations were replaced by a sum over $12 \times 12 \times 12$ and $8 \times 8 \times 1$ Monkhorst-Pack k-point meshes, respectively. The lattice constant of α-Sn was 6.4765 Å, which was obtained by the total energy minimization of a bulk α-Sn system described by a cubic unit cell for the diamond structure in the framework of the GGA+U method[29]. Then, the on-site Coulomb potential U = –2.5 eV was used for the calculation, as this value provides a better description of the electronic states in the bulk and slab geometries of α-Sn, such as the topological electronic structure around the band gap and the topological phase transition property[13, ,30,31].

*Tilted Weyl cones model.* Tilted Weyl cones are described by the Hamiltonian[24]

$$H_\chi = \chi v_F \boldsymbol{\sigma} \cdot \boldsymbol{k} + \boldsymbol{w}_\chi \cdot \boldsymbol{k}, \tag{2}$$

where $\chi = \pm 1$ is chirality of the Weyl nodes. The first term represents a linear dispersion with a Fermi velocity $v_F$, where an orbital degree of freedom (s or p orbital) of the Dirac electrons is regarded as a pseudo-spin and represented by a Pauli vector $\boldsymbol{\sigma}$. The second term represents a tilt $\boldsymbol{w}_\chi$ of the Weyl cones against the Fermi surface. When the Weyl cones are tilted oppositely to each other between the opposite chirality nodes ($\boldsymbol{w}_\chi = \chi \boldsymbol{w}$), the current density under an electric field $\boldsymbol{E}$ and a magnetic field $\boldsymbol{B}$ calculated by the semiclassical Boltzmann equation has a first-order component to $\boldsymbol{B}$[25] expressed as

$$\boldsymbol{j}^{(1)} \frac{4\pi}{15} \sigma_0 c_b [27(\boldsymbol{E}\cdot\boldsymbol{B})\boldsymbol{w} + 27(\boldsymbol{E}\cdot\boldsymbol{w})\boldsymbol{B} + 7(\boldsymbol{w}\cdot\boldsymbol{B})\boldsymbol{E}], \quad (3)$$

where $\sigma_0 = \frac{e^2 \tau \varepsilon_F}{4\pi^3 \hbar^3 v_F}$ and $c_b = \frac{e\hbar v_F^2}{\varepsilon_F}$ are constants. This component originates from non-zero Berry curvature and is non-vanishing only in the presence of the tilt term $\boldsymbol{w}$[25]. To compare the theoretical calculation with the experimental data, we set $\boldsymbol{E} = (E_x, 0, 0)$ and assume $\boldsymbol{w} = (w_x, w_y, 0)$ because the α-Sn layer has a two-dimensional Fermi surface. Then the relation between the calculated current density and the electric field is deduced as eq. (1) (ignoring the constants)

$$j^{(1)} \propto 61 w_x E B_x + 7 w_y E B_y.$$

*Magnetic sensor sensitivity:* The sensitivity of magnetic Hall sensors is typically quantified as $S = \frac{V_{Hall} B}{V_{in}}$ [mV/mT/V], where $V_{Hall}$ is the Hall voltage (in mV), $B$ is the magnetic field (in mT), and $V_{in}$ is the input voltage. Commercially available ultrahigh-sensitivity Hall sensors, which utilize high-mobility semiconductors such as GaAs, GaSb, and InSb, exhibit sensitivities in the range of 1.2–7 mV/mT/V [32]. For comparison, we estimate the sensitivity of a magnetic sensor based on the OMR observed in this study by substituting $V_{odd}$ (1T) and $V$ (0T) for $V_{Hall}$ and $V_{in}$, respectively. The resulting sensitivity is estimated to be 11.5 mV/mT/V—an order of magnitude higher than that of the best commercial ultrahigh-sensitivity Hall sensors.

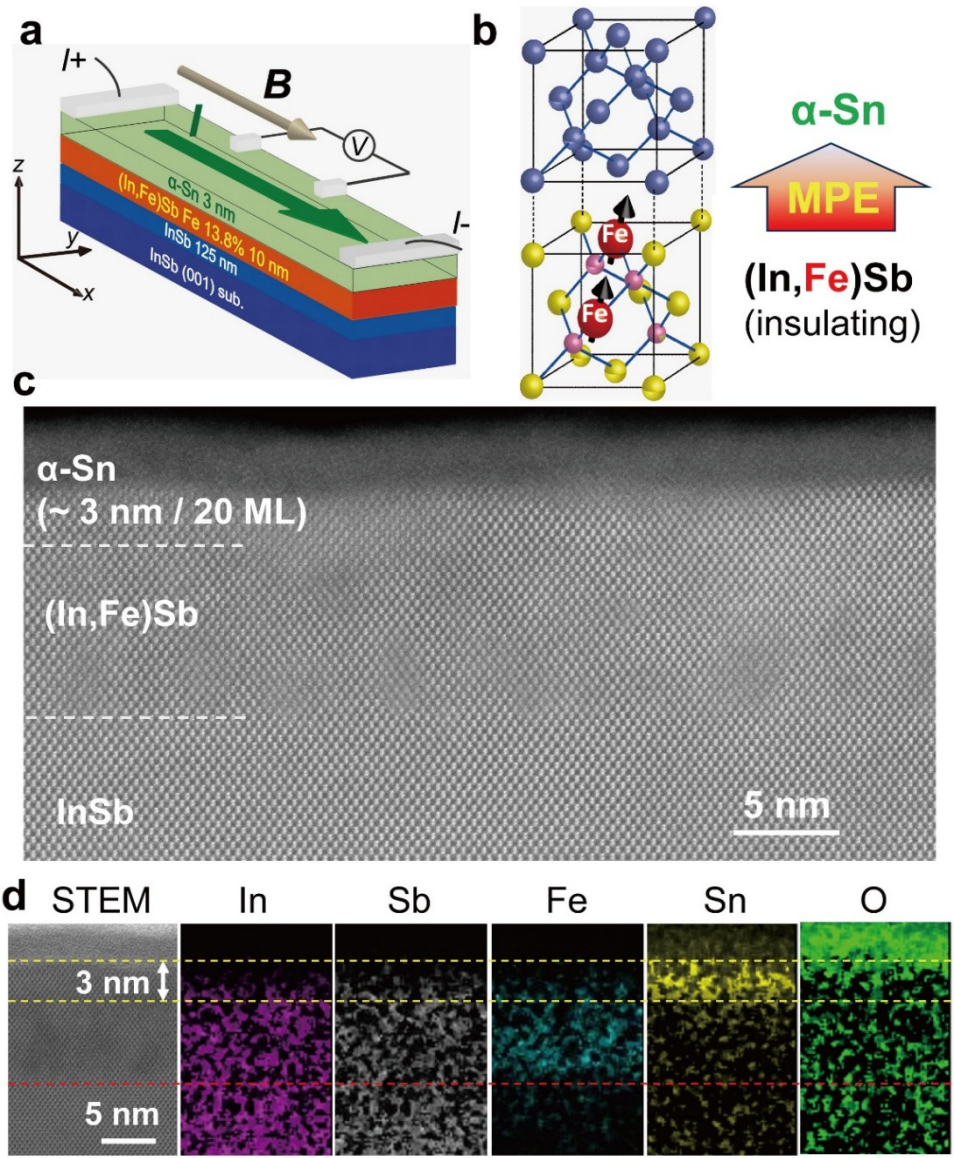

**Fig. 1 | Structural properties of the α-Sn/(In,Fe)Sb heterostructure sample. a**. Schematic illustration of the sample structure and magnetotransport measurement setup. **b**. Schematic crystal structure of α-Sn and FMS (In,Fe)Sb. Time-reversal symmetry (TRS) in α-Sn is broken due to the magnetic proximity effect (MPE) from (In,Fe)Sb. **c**. Cross-sectional scanning transmission electron microscopy (STEM) lattice image of the heterostructure taken along the [110] crystallographic axis of InSb. **d**. STEM lattice image and corresponding energy dispersive x-ray (EDX) elemental maps acquired on the same length scale. Yellow and red dashed lines indicate the interfaces between the layers as visual guides.

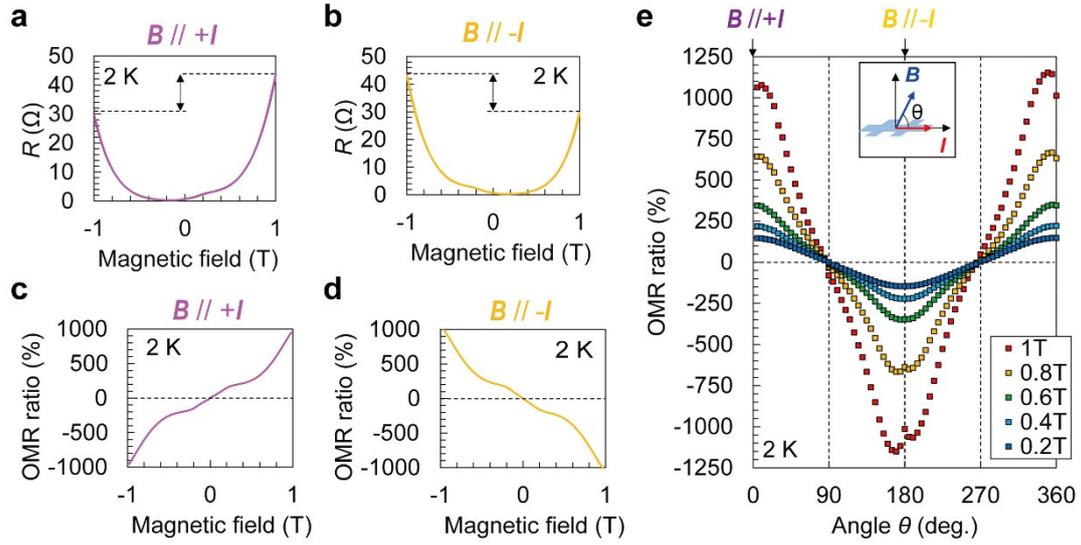

**Fig. 2 | Odd-parity magnetoresistance (OMR) observed in the α-Sn/(In,Fe)Sb bilayer heterostructure.** Longitudinal resistance measured under magnetic fields applied (**a**) parallel and (**b**) antiparallel to the direction of the current flow at 2 K. **c**, **d**. Corresponding OMR ratios, as defined in the main text, extracted from the data in **a** and **b**, respectively. **e**. Angular dependence of the OMR ratio as the magnetic field is rotated between the out-of-plane and in-plane directions (rotation in the *xz*-plane). All measurements were performed at various magnetic field strengths and at a temperature of 2 K.

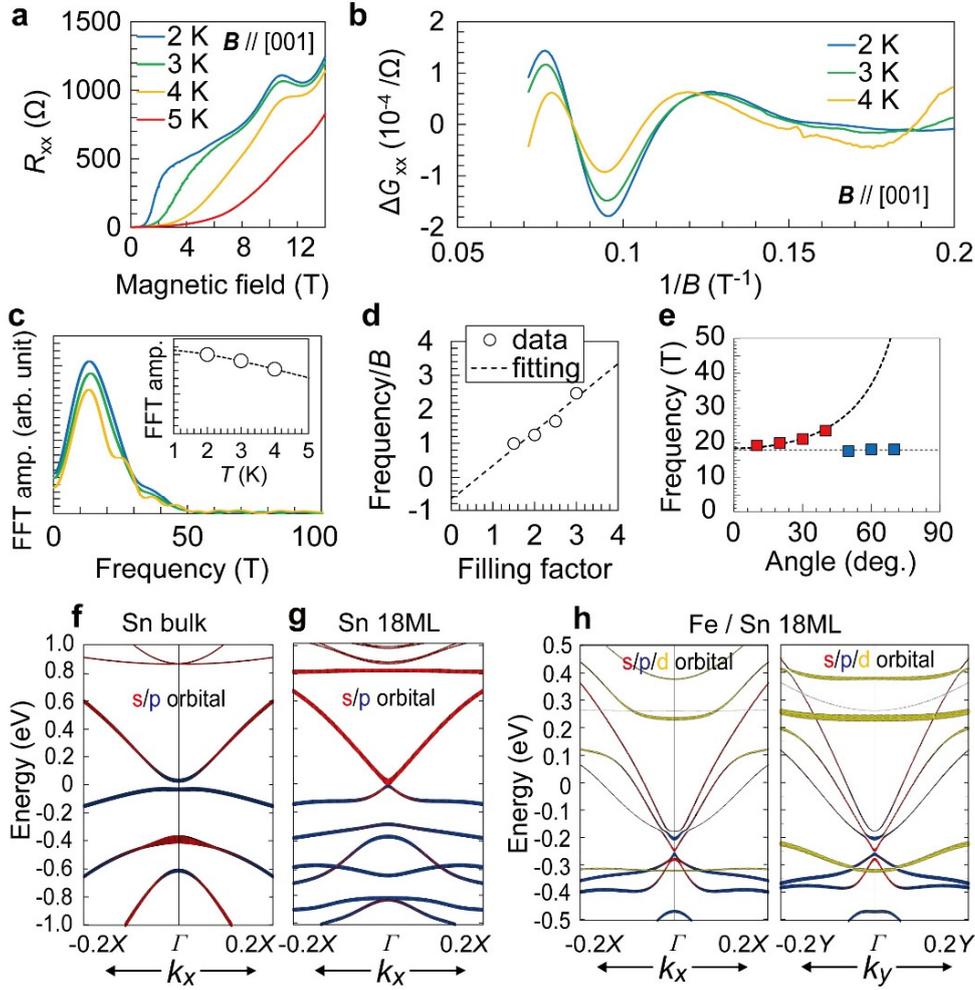

**Fig. 3 | Electronic band structure of the α-Sn/(In,Fe)Sb bilayer heterostructure. a**. Longitudinal resistances $R_{xx}$ measured under a perpendicular magnetic field at various temperatures. **b**. Shubnikov de Haas (SdH) oscillatory component $\Delta G_{xx}$ of the longitudinal conductance against the inverse magnetic field $1/B$. **c**. FFT spectra of the data in (**b**). Inset shows temperature dependence of the FFT peak intensity. **d**. Fan plot from the SdH oscillation. **e**. Angular ($\theta$) dependence of the peak frequency in FFT spectra of $dR_{yx}/dB$ when rotating the magnetic field from out-of-plane to in-plane directions. The dashed line is a $1/\cos\theta$ fit to the red points, and the dotted line is a constant fit to the blue points These red and blue points correspond to the 2D Fermi surface of α-Sn and the 3D Fermi surface of InSb, respectively. **f**, **g**, **h**. First principles calculation results of the band structures of (**f**) bulk α-Sn under tensile strain, (**g**) an 18-ML-thick α-Sn film grown on InSb, and (**h**) an 18-ML-thick α-Sn film with Fe atoms distributed on its surface along the $\Gamma$-$X$ and $\Gamma$-$Y$, respectively. Here, $X$ and $Y$ are on the Brillouin zone boundary in $\boldsymbol{k}$-space along the $k_x$ and $k_y$ directions, respectively (See Supplementary Fig. S4). The red, blue and yellow represent the states corresponding to s, p, and d-orbitals, respectively.

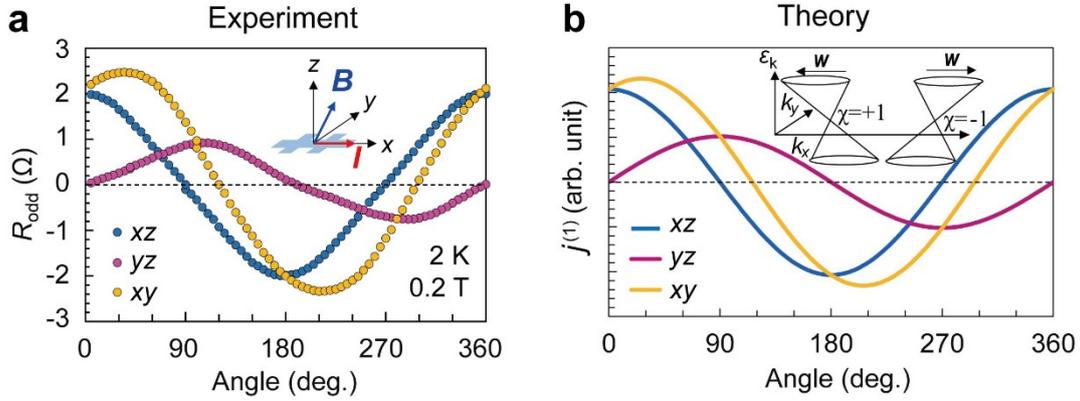

**Fig. 4 | Angular dependence of the OMR under magnetic field rotation. a**. Experimentally measured OMR as a function of magnetic field orientation. **b**. Theoretically calculated odd components of the current density based on Eq. (1), showing good agreement with the experimental data. The inset in (**a**) illustrates the coordinate system relative to the Hall bar structure and current direction. The inset in (**b**) shows a schematic representation of oppositely tilted Weyl cone dispersions used in the theoretical framework. The zero value of angle is defined when **B** is parallel to **I** (pointing to the *x* axis) in the *xz* and *xy* rotation data, and when **B** is parallel to the *y* axis in the *yz* rotation data.

# Supplementary Information

# Giant odd-parity magnetoresistance from proximity-induced topological states

Tomoki Hotta[1], Le Duc Anh[1,2*], Takahiro Chiba[3,4], Yohei Kota[5], and Masaaki Tanaka[1,2,6*]

[1]Department of Electrical Engineering and Information Systems, The University of Tokyo
[2]Center for Spintronics Research Network (CSRN), The University of Tokyo
[3] Department of Information Science and Technology, Graduate School of Science and Engineering, Yamagata University, Japan
[4]Department of Applied Physics, Graduate School of Engineering, Tohoku University
[5]National Institute of Technology, Fukushima College
[6]Institute for Nano Quantum Information Electronics (NanoQuine)

## Supplementary Table and Figures

**Supplementary Table S1**| Comparison of the OMR values, defined as $[R(B) - R(-B)] / [2 \times R(0\ \mathrm{T})]$, reported so far.

| Material | Maximum OMR (%) | $B$ (T) | Reference |
|---|---|---|---|
| $Eu_2Ir_2O_7$ | 0.4% | 9 | [S1] |
| $SmCo_5$ | 0.05% | 0.5 | [S2] |
| InAs/(Ga,Fe)Sb | 13.5% | 10 | [S3] |
| bilayer graphene/$Cr_2Ge_2Te_6$ | 40% | 14 | [S4] |
| α-Sn/(In,Fe)Sb | **1150%** | **1** | **This work** |

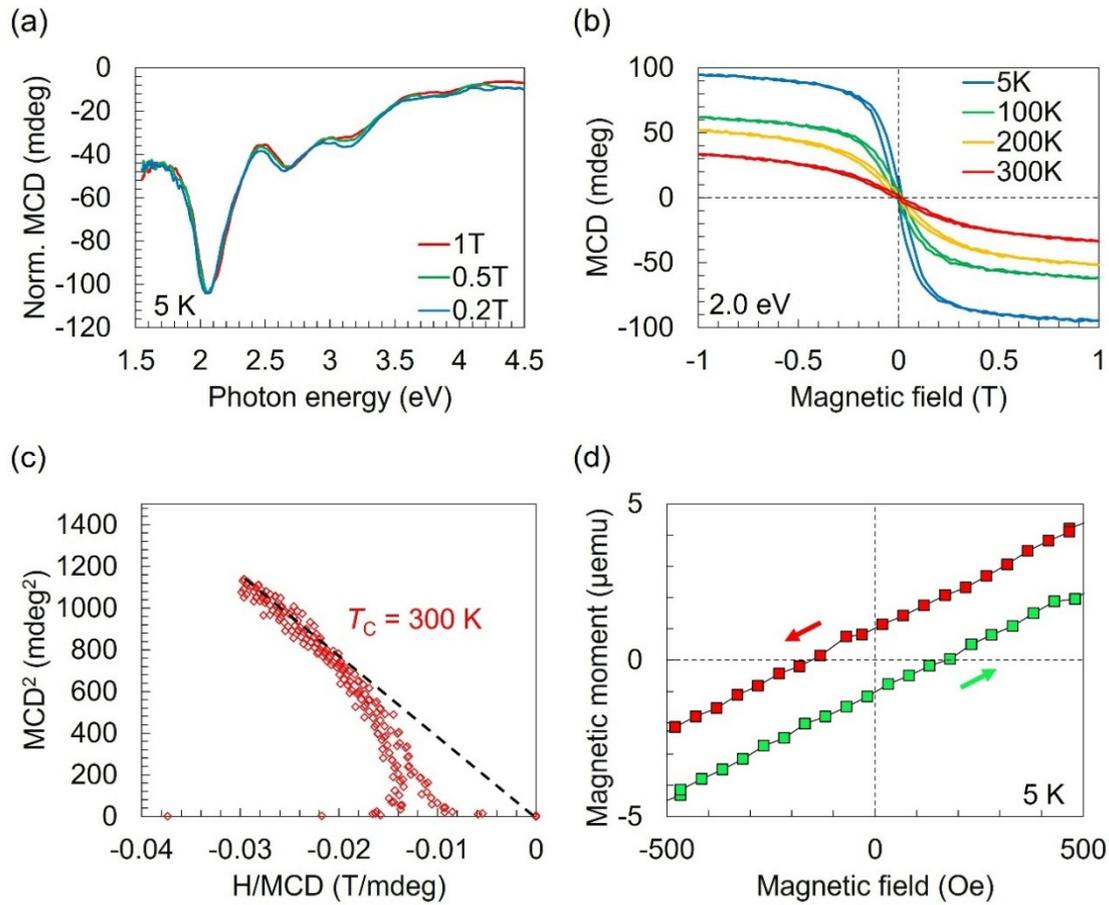

**Supplementary Fig. S1 | Magnetic properties of the (In,Fe)Sb layer.** (a) Reflection magnetic circular dichroism (MCD) spectra measured at 5 K under various perpendicular magnetic fields. MCD intensities at 0.5 T and 0.2 T are normalized to the spectrum at 1 T. (b) MCD intensity – magnetic field (*MCD – H*) curves measured at various temperatures. The MCD intensity was at taken at 2.0 eV. (c) Arrott plot derived from the *MCD – H* curve at 300 K, used to estimate the Curie temperature. (d) Magnetization – magnetic field (*M – H*) curve measured by SQUID magnetometry at 5 K.

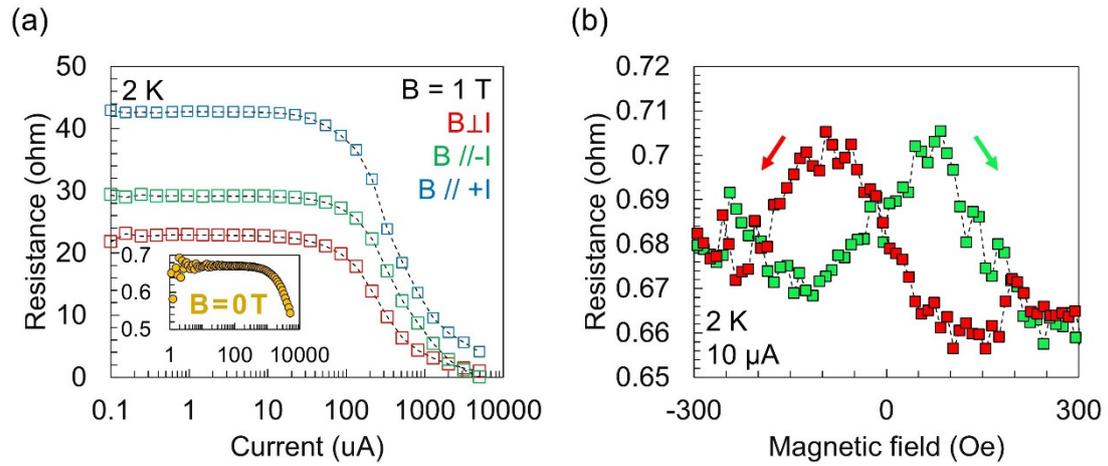

**Supplementary Fig. S2 | Current and magnetic field dependences of the longitudinal resistance at 2 K.** (a) Current dependence of the longitudinal resistance measured at 2 K. Red, green, and blue data points correspond to the measurements under a 1 T magnetic field applied perpendicular to the film plane (***B*** ⊥***I***), oriented antiparallel (***B***//–***I***), and parallel (***B***//+***I***) to the current direction, respectively. The inset shows data (yellow points) taken under zero magnetic field. (b) Magnetic field dependence of the longitudinal resistance at 2 K under a perpendicular magnetic field. The peak positions (±100 Oe) correspond to the coercive field in the *M – H* curve of Fig. S1(d).

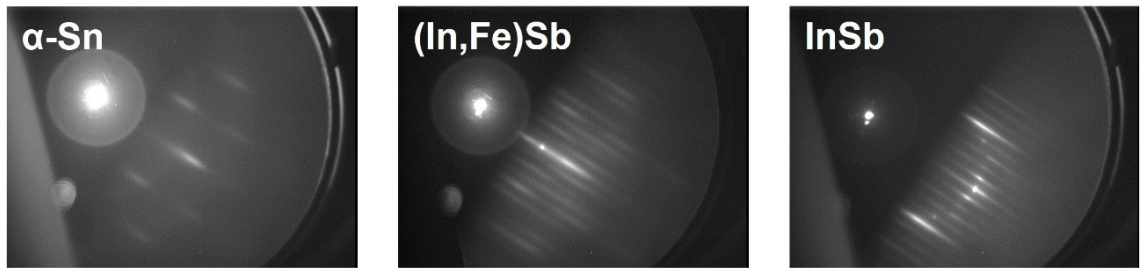

**Supplementary Fig. S3 |** RHEED pattern of each layer during the MBE growth along the [110] direction.

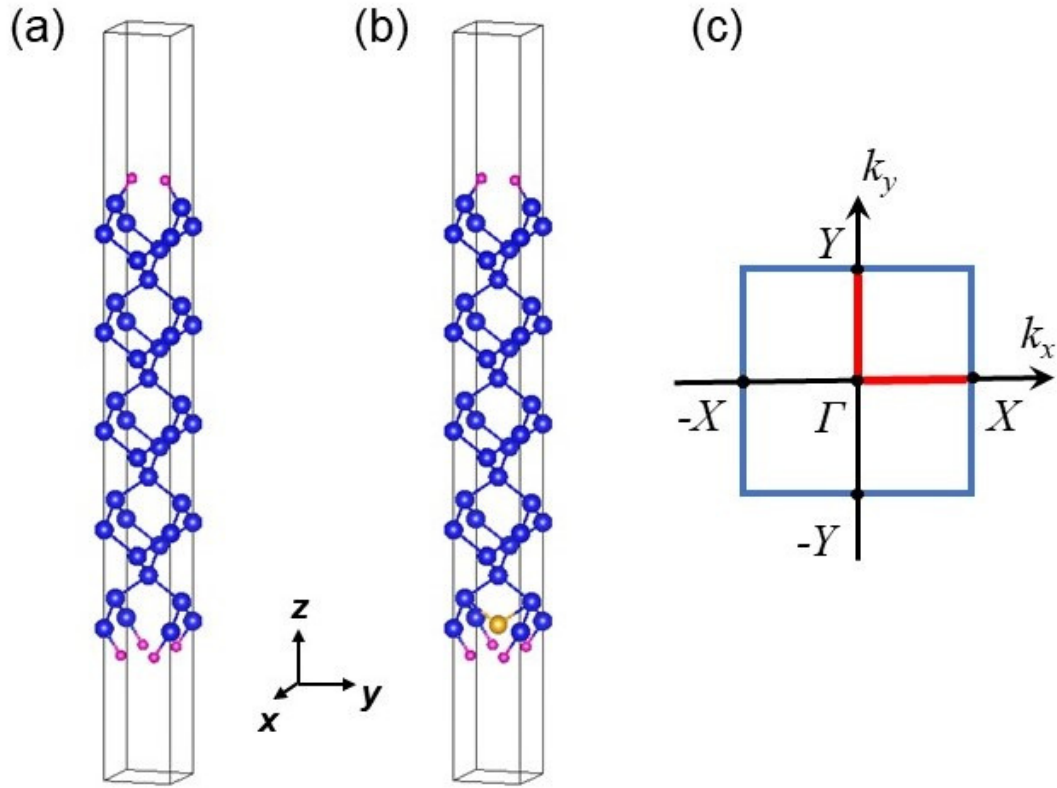

**Supplementary Fig. S4 | Crystal structures used in slab calculations.** (a) Slab geometry of α-Sn/vacuum. (b) Slab geometry of α-Sn/Fe along the [001] direction. Sn, Fe, and H atoms are represented by blue, yellow, and pink spheres, respectively. (c) Definition of the two-dimensional Brillouin zone in our calculation. Here, $X = (\pi/a^*, 0)$ and $Y = (0, \pi/a^*)$ are on the Brillouin zone boundary in the $k_z = 0$ plane. $a^* = a/\sqrt{2}$ is the in-plane lattice constant of the tetragonal unit cell of α-Sn in our calculations, where $a$ is the lattice constant of the diamond-type unit cell of α-Sn.

**Supplementary note 1**

We characterized the magnetic properties of our heterostructure sample using magnetic circular dichroism (MCD) spectroscopy and superconducting quantum interference device (SQUID) magnetometry. Supplementary Fig. S1(a) presents the MCD spectra measured at 5 K under magnetic fields of 1 T, 0.5 T, and 0.2 T, applied perpendicular to the film plane. The spectra obtained at 0.5 T and 0.2 T are normalized to that at 1 T. All the normalized spectra are almost perfectly overlapped, featuring a negative peak at 2.0 eV ($E_1$) and a positive peak at 2.5 eV ($E_1 + \Delta_1$)—a spectral signature characteristic of the intrinsic ferromagnetism in (In,Fe)Sb [S5]. The magnetic field dependence of the MCD intensity (*MCD–H* curves), measured at a photon energy of 2.0 eV [Supplementary Fig. S1(b)], shows clear ferromagnetic hysteresis. From the Arrott plot analysis[S6], the Curie temperature ($T_C$) is estimated to be 300 K [Supplementary Fig. S1(c)]. SQUID magnetometry at 5 K [Supplementary Fig. S1(d)] reveals a coercive field of approximately 100 Oe, further confirming the ferromagnetic nature of the (In,Fe)Sb layer.

**Supplementary note 2**

Supplementary Figure S2(a) shows the current (*I*) dependence of the longitudinal resistance under a magnetic field of 1 T. For currents below 10 μA, the resistance remains constant, indicating stable conduction behavior. However, when the current exceeds 10 μA, the resistance decreases monotonically with increasing the current. This behavior is likely attributed to a change in the current distribution. At low current ($I < 10$ μA), conduction predominantly occurs within the α-Sn layer due to its semi-metallic nature. It is worth noting that the (In,Fe)Sb layer is insulating at low temperatures because of carrier localization induced by Fe 3d states[S5]. The observed resistance decrease at higher currents ($I > 10$ μA) may be due to current leakage into the underlying InSb buffer layer, possibly penetrating through the nominally insulating (In,Fe)Sb.

All the transport measurements in this study were performed at $I = 10$ μA, ensuring that the current primarily flows through the α-Sn layer. Under this condition, we also observe clear hysteresis in the magnetoresistance at 2 K [Supplementary Fig. S2(b)], with peaks at ±100 Oe corresponding to the coercive field of approximately 100 Oe [Supplementary Fig. S1(d)]. This hysteresis indicates that the α-Sn layer becomes magnetized via the adjacent ferromagnetic (In,Fe)Sb layer. Such a phenomenon, where the magnetic order of one material influences an adjacent layer through quantum mechanical coupling, is known as the magnetic proximity effect.

## Supplementary note 3

Here, we comment on the validity of comparing the resistance and current in the context of our measurements and theoretical model. In the experiment, a constant direct current $I_{DC}$ is applied, and the voltage change upon magnetic field reversal is detected as the OMR signal: $V_{odd}^{exp.} = R_{odd}^{exp.} \times I_{DC}$. On the other hand, semiclassical transport theory—where the net current arises from a shift of a rigid Fermi surface—predicts a voltage signal proportional to the current density: $V_{odd}^{calc.} \propto j_{odd}^{calc.}$. Assuming that the experimentally observed voltage equals the theoretically calculated one, $V_{odd}^{exp.} = V_{odd}^{calc.}$, it follows that: $R_{odd}^{exp.} \propto j_{odd}^{calc.}$. This proportionality justifies the comparison between the experimentally measured OMR ratio and the theoretically calculated odd component of the current density.

## Supplementary note 4

For both the α-Sn/vacuum and α-Sn/Fe slab geometries, we used the layered tetragonal unit cell based on the diamond-type crystal structure deposited along the [001] direction, as shown in Supplementary Fig. S4. The 18 ML slab geometry was adopted for the computation of the topological surface state (TSS) of strained α-Sn. Both edges of the slab were terminated with H atoms to remove dangling bonds. On the other hand, for the computation of a strained α-Sn/Fe system, it was adopted that the 18 ML slab geometry in which a single Fe atom is embedded as a dopant on the lowest Sn layer. For both of the slab geometries, the surfaces were separated by a vacuum layer with a thickness of 20 Å. To calculate the strain effect on the topological electronic states, we considered a biaxial in-plane compressive strain of –0.76%[S7]. Fixing the in-plane lattice constant, structural optimization of atomic position was performed.

## Supplementary References